\documentclass{elsart}
\usepackage{epsfig,float,amssymb}
\journal{Astroparticle Physics}

\begin{document}

\newcommand{\light}{\operatorname{c}}

\newcommand{\aeta}{{\em Astron. Astrophys.}}
\newcommand{\astrophysj}{{\em Astrophys. Jour.}}
\newcommand{\mnras}{{\em Mon. Not. R. Ast. Soc.}}
\newcommand{\nuclphysa}{{\em Nucl. Phys. A}}
\newcommand{\physrevl}{{\em Phys. Rev. Lett.}}
\newcommand{\apjlet}{{\em Astrophys. Jour. Lett.}}
\newcommand{\prd}{{\em Phys. Rev. D}}
\newcommand{\apj}{{\em Astrophys. J.}}
\newcommand{\pl}{{\em Phys. Lett.}}

\begin{frontmatter}

\title{Detection of a close supernova gravitational wave burst in a network of interferometers, neutrino and optical detectors}

\author[lal]{Nicolas Arnaud\corauthref{cor}},
\corauth[cor]{Corresponding author.}
\ead{narnaud@lal.in2p3.fr}
\author[lal]{Matteo Barsuglia},
\author[lal]{Marie-Anne Bizouard},
\author[lal]{Violette Brisson},
\author[lal]{Fabien Cavalier},
\author[lal]{Michel Davier},
\author[lal]{Patrice Hello},
\author[lal]{Stephane Kreckelbergh},
\author[lal]{Edward K. Porter}

\address[lal]{Laboratoire de l'Acc\'el\'erateur Lin\'eaire, CNRS-IN2P3 and Universit\'e Paris Sud,\\B.P. 34, B\^atiment 200, Campus d'Orsay, 91898 Orsay Cedex (France)\protect\\}

\begin{abstract}

Trying to detect the gravitational wave (GW) signal emitted by a type II supernova is a main challenge for the GW community. Indeed, the corresponding waveform is not accurately modeled as the supernova physics is very complex; in addition, all the existing numerical simulations agree on the weakness of the GW emission, thus restraining the number of sources potentially detectable. Consequently, triggering the GW signal with a confidence level high enough to conclude directly to a detection is very difficult, even with the use of a network of interferometric detectors. On the other hand, one can hope to take benefit from the neutrino and optical emissions associated to the supernova explosion, in order to discover and study GW radiation in an event already detected independently. This article aims at presenting some realistic scenarios for the search of the supernova GW bursts, based on the present knowledge of the emitted signals and on the results of network data analysis simulations. Both the direct search and the confirmation of the supernova event are considered. In addition, some physical studies following the discovery of a supernova GW emission are also mentioned: from the absolute neutrino mass to the supernova physics or the black hole signature, the potential spectrum of discoveries is wide.

\end{abstract}


\end{frontmatter}


\section{Introduction}

Type II supernovae have been considered for years as one of the most promising sources of gravitational waves (GW); indeed, in the 60's-70's, the first GW detectors -- resonant bars -- had their resonant frequencies tuned around one kHz, in order to look for the typical frequencies expected for these GW signals. However, the amount of energy $E_{\mathrm{GW}}$ released through GW during the supernova explosion was overestimated by orders of magnitude at that time. Later, a clear upper bound for this quantity was established: $\mathrm{E}_{\mathrm{GW}} \lesssim 10^{-6} M_{\odot}c^2$ \cite{BM}, with $M_\odot$ being the solar mass. This value was confirmed by other numerical analyzes, both with Newtonian simulations \cite{ZM} and with methods including a relativistic treatment of the rotational core collapse \cite{DFM}. The latter study provides an even more restrictive limit at a few $10^{-7} M_{\odot}c^2$. Consequently, detecting a GW supernova signal with the first generation of GW interferometric detectors currently being developed -- GEO600 \cite{geo}, the two LIGO interferometers \cite{ligo}, TAMA300 \cite{tama}, Virgo \cite{virgo} -- or foreseen -- ACIGA \cite{aciga} -- appears very challenging.

The bounce following the stellar collapse is expected to last only a few milliseconds, during which most of the power radiated in GW is emitted. The corresponding GW waveforms cannot be well predicted as the physics of this phenomenon is too complex to be studied in details. Yet, more and more signal models have been computed by various groups (for a survey, see e.g. \cite{DFM} and references therein), and the situation seems now converging: new simulations confirm the previous ones, and even if the computed amplitudes are still spread over a large range, first conclusions with direct implications on data analysis can be drawn, as shown in Section \ref{section:SN_physics}. In the same section, the two other main emissions occurring during a supernova explosion (neutrinos and photons) are recalled as they will be shown to facilitate the GW detection in the following. Indeed, if a direct search of the GW signal is unfruitful, alternative methods taking advantage from these other emissions can be used, as neutrinos would bring informations on the timing of the supernova (and marginally on the source location), while the optical signature would provide the source position in the sky.

Section \ref{section:GW_detection} deals with the search of the supernova GW signal and presents various strategies, depending on the goals of the analysis -- either a direct and independent detection or the confirmation of an already-known event -- and on the available informations

Supernova GW signals belong to the generic category of bursts, short (a few ms) signals poorly-modeled. Detecting these GW bursts always requires the use of a network of interferometers, in order to discriminate between a real GW signal -- 'present' in some sense in all detectors -- and transient noises regularly occurring in interferometers, but limited to a single instrument at a given time. In Ref. \cite{arnaud_burst_2,arnaud_burst_3}, a simulation of the detection process in a network of interferometers was introduced, and then used to compare quantitatively the two main data analysis strategies: the coincidence and the coherent approach. Here, we complete this work by focusing on the case where the supernova has been detected independently of the GW detectors, through its neutrino emission or its optical signature. These additional informations can reduce considerably the difficulties inherent in searching for GW signals and allow for instance the use of an additional filtering method, a logical 'OR' of all the network component outputs.

This 'confirmation' scenario is not only realistic given the current knowledge of the different supernova emissions, but also really attractive: confirming that a supernova emits also GW would be a major discovery by itself, if not the first direct proof of the existence of GW. In this spirit, Section \ref{section:physics} recalls the main physical outputs one can expect from the detection of the three signals emitted by a supernova: GW, neutrinos, and the optical flare. For instance, the time delay between the GW and the neutrino emission gives informations on the neutrino absolute masses \cite{MD_neutrino}. One can also test models of the supernova explosion, or identify a delayed black hole formation from its characteristic neutrino and GW signatures.

For completeness, the main hypotheses of the network data analysis Monte-Carlo simulations and some notations used in the core of this article are summarized in Appendix \ref{section:itf_network_simulation}.

\section{Type II supernova physics overview}
\label{section:SN_physics}

A type II supernova is a violent event ending the life of massive enough -- above 8 $M_\odot$ -- stars once they have burnt all their light elements; it releases a huge amount of energy, about $3 \times 10^{46}$~J in total. Less than 1\% is dissipated in kinetic energy or optical emission, and almost all the energy is emitted through neutrinos, first during a millisecond burst of $\nu_e$, associated to the neutronization phase following the core collapse, and then via a longer and dominantly thermal emission equally distributed between all neutrino flavours. Such explosions are expected to emit GW too, almost in coincidence with the core bounce, but its amplitude is so small that it has not been detected yet.

In this section, the main features of the three supernova signals used in the analyzes presented later in this article are briefly recalled, with a particular emphasis on the GW part. Before that, the next paragraph summarizes the present status on the supernova rate: as only close supernovae are expected to be detectable, estimating their occurrence rate is also important.

\subsection{Supernova rate}
\label{subsection:SN_rate}

The supernova occurrence per galaxy depends on its type: for instance, type II supernovae have young progenitors and so can only occur either in new galaxies or in star formation areas. They are absent in (old) elliptic galaxies, whereas they are more common in spiral or irregular ones. Table~1 shows the supernova rates versus the galaxy type extracted from Ref. \cite{Cappellaro}; the numbers are expressed in SNu\footnote{1 SNu = 1 supernova per century and per $10^{10} \, \mathrm{L}_\odot$ with the luminosity expressed in the Blue band (wavelength $\sim \, 0.45 \, \mu$m)}. The average value is about $\left( 0.68 \, \pm \, 0.20 \right)$ SNu.

Assuming these rates, the Tully catalog of nearby galaxies \cite{Tully} can be used to predict the expected number of supernovae per year up to a given distance. The catalog contains 2367 galaxies up to 40 Mpc. If the Blue luminosity of a galaxy is not known, it can be estimated from the catalog data themselves, by using either the mean mass/luminosity ratio for the corresponding type of galaxy, or its hydrogen mass value.

Similar studies have been already presented, for instance in Ref. \cite{Paturel} where a much larger catalog of galaxies and higher supernova rates -- thus leading to more optimistic predictions -- have been used. As shown later in Section \ref{subsection:GW_emission}, only close supernovae are expected to be visible in interferometers; therefore, the Tully catalog is enough to estimate the detectable rate of events and no significant correction taking into account galaxies with apparent magnitudes too low to be observed is required.

Figure~1 gives the number of supernovae per year versus the distance. The steps in the various curves at 17 Mpc correspond to the Virgo cluster: in order to reach a rate of a few events per year, one should be sensitive up to this distance. The shapes of the various curves also change when the Virgo cluster is reached: beyond, they show a power-law behavior while for closer distances the distribution is irregular. Yet, as noticed earlier e.g. by Ref. \cite{Paturel}, the power one can extract from a fit of the supernova rate beyond the Virgo cluster is close to 2.2, i.e. below the value of 3 which would correspond to a uniform density of sources.

Aborted or silent supernovae are not included in the surveys as they have not been optically seen. Yet, one can infer from neutrino experiment results that their rates in the Milky Way are not much higher than those of 'classical' supernovae, as no such signal has been recorded in about a decade of running for these detectors. However, direct black hole formations without neutrino emission may have also occurred; as such events could only be seen in GW detectors, their rate is presently unknown.

To conclude, one can estimate the Galactic supernova rate. The Milky Way type is Sbc and its luminosity is about $2 \, 10^{10} \, \mathrm{L}_\odot$. Therefore, the expected number of supernovae is 2.4 per century ($\sim$ 1 in 40 years), among which 70\% are of type II. A different analysis presented in \cite{Panagia} assuming that only 10\% of the supernovae are seen visually gives a more interesting rate of 1 in 20 years, as 9 events have been recorded for the two past millennia. Nevertheless, the supernova galactic rate remains small; yet, with some luck, few such events could be recorded during the lifetime of current GW antennas. The fact that this possibility may be unique gives even more motivation to be able to track it in the best possible way.

\subsection{Known supernova signatures}
\label{subsection:nu_opt}

\subsubsection{Neutrinos}

For a review focused on this topic, one can see for instance Ref. \cite{Cei} and references therein. As mentioned above, most of the energy released during a supernova is emitted through neutrinos. First, there is a short -- $\sigma_{\mathrm{flash}} \sim ( 2.3 \pm 0.3 )$ ms -- but intense (a few $10^{44}$~J, about 1\% of the total amount of energy carried away by neutrinos) neutronization burst of electron neutrinos, which increases rapidly before decaying exponentially.

After the bounce, neutrinos of all flavors are thermally produced as $\nu\bar{\nu}$ pairs; this cooling emission is supposed to last a few seconds, and dominates the phenomenon after the $\nu_{\mathrm{e}}$ burst. The energy is almost equally distributed between the different neutrino types: $5 \times 10^{45}$~J per flavor of neutrino or anti-neutrino. Typical mean energies for neutronization and thermal neutrinos are $E_\nu \sim 10-20$~MeV.

Assuming the supernova to be located in the Galaxy or at least in the Local Group -- the only region potentially visible by the first generation of interferometric detectors given the expected amplitude of the GW signals, as shown in Section \ref{subsection:GW_emission} --, the different flavor emissions will certainly be detected by the current generation of underground neutrino detectors: SuperKamiokande \cite{SuperK}, SNO \cite{SNO} or LVD \cite{LVD}.

Indeed, Table~2 shows the number of neutrinos expected in the different detectors previously mentioned; the numbers are scaled for a source at 10 kpc, a value close to the Galactic center distance (8.5~kpc) and to the mean distance for a Galactic source (around 11~kpc). It is clear that enough neutrinos would be detected, providing a clear signature of the event and an accurate timing. Since the neutrino flux varies as $D^{-2}$, supernova detection is not expected much beyond 200 kpc. It is worth noting that a neutrino signal can be used alone to locate the supernova in the sky but its analysis provides at best a rough information on the source location \cite{beacom_vogel}.

\subsubsection{Optical flare}

All these neutrino experiments belong to the SNEWS network \cite{SNEWS} aiming at delivering a fast alert (within a few tens of minutes) when some of its components are triggered by an unexpected 'high' flux of neutrinos. Indeed, this is fast enough to activate optical telescopes looking for the light flare, occurring with a time delay of several hours. Provided that the light emission is neither obscured by some interstellar dust, nor hidden by a too bright neighborhood, this signature can be visible over cosmological distances in efficient telescopes: its luminosity can be as high as the host galaxy (about few $10^{35}$~W).

\subsection{GW emission}
\label{subsection:GW_emission}

For reasons explained above, only numerical results are available for the GW waveforms from core collapses. Even if the generated signals have some common features -- short timescales, one or more main peaks, possibly damped oscillations -- the simulation results are model-dependent, both for what concern their shapes and their amplitudes. Some quantities or concepts introduced in this section are presented in more details in Appendix \ref{section:itf_network_simulation}.

\subsubsection{Maximal amplitude comparison}

Figure~2 presents the maximum amplitudes computed either in simulations \cite{BM,ZM,DFM,GW_simul_1,GW_simul_3,GW_simul_4} or by indirect studies of pulsar velocities \cite{pulsar_velocity_1,pulsar_velocity_2}. Apart one very optimistic prediction \cite{GW_simul_3} based on an analogy with the collapse of molecular cloud into a proto-star, maximum amplitudes at the Virgo cluster distance are all below $10^{-22}$, too weak to expect a detection of such distant supernovae, with present detectors. On the other hand, Galactic supernovae should be clearly visible.

One can also note that the GW maximal amplitudes are spread over three orders of magnitudes between the different simulation groups and, even within the same numerical group -- such as \cite{ZM} or \cite{DFM} whose waveforms appear in a medium position on the plot --, the spread can reach two orders of magnitude. Yet, this does not modify the general picture concerning the supernova GW signal detectability.

\subsubsection{Energy comparison}
\label{subsubsection:E_comp}

Indeed, a more precise way to compare simulated waveforms consists in using the energy $E_{\mathrm{GW}}$ emitted through GW instead of the maximum amplitude. Following Thorne \cite{thorne87}, one can link this quantity to the GW amplitude $h(t)$:

\begin{equation}
E_{\mathrm{GW}} \; = \; \frac{ 2 \, c^3 \, \pi^2 \, D^2 }{ G } \; \int_0^{+\infty} \, df \, f^2 \, \left| \tilde{h}(f) \right|^2
\label{eq:E_GW_1}
\end{equation}
where $\tilde{h}$ is the Fourier transform of $h(t)$, scaling as $D^{-1}$, with $D$ the source distance. Assuming a given waveform, one can infer the dependence of the signal-to-noise ratio (SNR) in the GW energy $E_{\mathrm{GW}}$.

For a Gaussian GW burst of half-width $\omega$ and peak amplitude $h_{\mathrm{peak}}$ -- as defined in Appendix \ref{section:itf_network_simulation} --, Eq. (\ref{eq:E_GW_1}) becomes

\begin{equation}
E_{\mathrm{GW}} \; = \; \frac{ c^3 \; \sqrt{\pi} \; h_{\mathrm{peak}}^2 }{ 8 \; G \; \omega }
\label{eq:E_GW_2}
\end{equation}

On the other hand, the optimal SNR $\rho_{\mathrm{max}}$ -- see Appendix \ref{section:itf_network_simulation} -- is given by

\begin{equation}
\rho_{\mathrm{max}}^2 \; = \; \frac{ \sqrt{\pi} \; f_{\mathrm{samp}} \; \omega \; h_{\mathrm{peak}}^2 }{ \sigma_{\mathrm{noise}}^2 \; D^2 } 
\label{eq:rho_max}
\end{equation}
with $f_{\mathrm{samp}}$ being the detector sampling frequency, and $\sigma_{\mathrm{noise}}$ the detector white noise RMS.

As simulated supernova GW bursts exhibit shapes more complicated than a single peak, only part of the energy can be associated to this component of the signal: let $\kappa_{\mathrm{eff}}$ be the corresponding reduction factor (in GW amplitude). Merging together Eq. (\ref{eq:E_GW_2}) and (\ref{eq:rho_max}) gives finally

\begin{equation}
\rho_{\mathrm{max}} \; \approx \; 12.7 \; \left( \frac{ \kappa_{\mathrm{eff}} }{ 0.5 } \right) \; \left( \frac{ \omega }{ 1 \, \mathrm{ms} } \right) \; \left( \frac{ 4 \times 10^{-21} }{ \sigma_{\mathrm{noise}} } \right) \; \left( \frac{ 8.5 \, \mathrm{kpc} }{ D } \right) \; \sqrt{ \frac{ E_{\mathrm{GW}} }{ 10^{-8} \, M_\odot \, c^2 } }
\label{eq:rho_vs_E}
\end{equation}
The value $\sigma_{\mathrm{noise}} = 4 \times 10^{-21}$ corresponds to the RMS of a white Gaussian noise, with an amplitude spectrum density $4 \times 10^{-23} / \sqrt{Hz}$ -- the minimum of the nominal sensitivity of the Virgo interferometer --, and sampled at a frequency $f_{\mathrm{samp}}=20$~kHz.

Figure~3 uses this model to show how the optimal SNR $\rho_{\mathrm{max}}$ scales with the distance, given the numerical estimations of the GW radiated energy. As already stated, detecting a Galactic supernova seems realistic from this plot, while this possibility is completely ruled out for events in the Virgo cluster. Even the distance of 1 Mpc appears to be beyond the interferometer sensitivity, but the Magellanic clouds (distant of few tens of kpc) should belong to the sensitive volume.

\subsubsection{Summary}

The main conclusions drawn from Figures~2 and~3 are twofold. On the one hand, detecting a supernova further than a few hundred kpc appears impossible with the currently planned sensitivities of the interferometers if the computed GW amplitudes are realistic. On the other hand, supernovae in the Milky Way are likely to be seen. This validates our choice of focusing our attention on Galactic events for which neutrino detectors are also well-matched.

A last point worth being mentioned is that most of the simulations presented in this section assume that the core collapse is axisymmetrical; this hypothesis is a consequence of the present computing limitations. However, the larger the asymmetry, the stronger the GW emission and so current simulations could well underestimate the real supernova signal. Yet, this effect cannot be estimated properly, and the results of Ref. \cite{BM} assuming no such symmetry do not show a significant improvement in the GW amplitudes. Therefore, we prefer to be conservative in this study, by considering that current simulations describe adequately the supernova GW magnitude.

\subsection{Delay between neutrino and GW emissions}
\label{subsection:delay_nu_GW}

Simulations of core collapses show a very strong correlation between neutrino and GW signals. This section briefly summarizes the current knowledge on their relative timing, and the experimental precision one can expect for each of the two emissions. 

\subsubsection{Neutrino and GW signal timing accuracy}

The statistical precision on the arrival time of the electron neutrino flash depends on the number $N_e$ of $\nu_{\mathrm{e}}$ detected. The RMS of the Poissonian timing error is given by

\begin{equation}
\sigma_{\nu} \; = \; \frac{ \sigma_{\mathrm{flash}} }{ \sqrt{ N_e } }
\label{eq:nu_stat}
\end{equation}
Even if the $\nu_{\mathrm{e}}$'s have oscillated, a precision on the order of 1 ms can still be achieved from neutral current interactions in SNO \cite{MD_neutrino}.

For the GW signal, the timing accuracy strongly depends on the filtering method chosen. Various Monte-Carlo studies have already been performed on this topic \cite{arnaud_burst_2,arnaud_burst_3,hello_timing}, in order to compute the systematics associated to these algorithms and to estimate their statistical precision in a single detector. The results of these simulations are contrasted: on the one hand, locating simulated waveforms with complicated shapes is very difficult; on the other hand, Gaussian-like peaks can be well timed, as soon as they are detected with a SNR $\rho$ high enough.

The RMS of the statistical timing error $\sigma_{\mathrm{GW}}$ can be parameterized by the following generic formula:

\begin{equation}
\frac{ \sigma_{\mathrm{GW}} }{ \omega } \; = \; K \; \left( \frac{ 10 }{ \rho } \right)^\alpha
\label{eq:GW_stat}
\end{equation} 
with $\omega$ being the Gaussian half-width and the pair $(K,\alpha)$ being specific for a filter. Table~3 presents the values of these parameters for three different methods briefly described below -- see also Appendix \ref{section:itf_network_simulation}. In all cases, the precision achieved is well below the signal width and decreases quickly with the filter output $\rho$.

\begin{itemize}
\item The Wiener filter, optimal for known waveform, thus giving the best results.
\item The mean filter \cite{hello_timing} which monitors the average value of the data in the analysis window, shifted in time. Despite its simplicity, this filter has been found to be quite powerful for the search of poorly modeled GW bursts.
\item ALF \cite{pradier}, a non linear algorithm based on a linear fit of the data whose outputs (a slope and an offset) are optimally combined. It gives currently the best mean detection efficiency on the Zwerger-M\"uller catalog \cite{ZM}.
\end{itemize}

\subsubsection{GW and neutrino relative timing}

As the start of the neutronization burst corresponds to the moment when the shock wave induced by the core bounce reaches the neutrinosphere, one can infer a relation between the $\nu_{\mathrm{e}}$ peak and the bounce timing. Indeed, from numerical simulations \cite{ZM,DFM}, one obtains:

\begin{equation}
\Delta t_{\nu_{\mathrm{e}} \, , \, \mathrm{bounce}} \; \approx \; (3.5 \, \pm \, 0.5) \; \mathrm{ms}
\label{eq:nu_bounce}
\end{equation}
This value can be simply estimated by noticing that the neutrino sphere radius is about 100~km and that the shock wave propagates with a speed of order $c/10$. In addition, numerical simulations of supernovae show a clear correlation between the maximum of the GW and the core bounce. Using the 78 signals of the Zwerger-M\"uller catalog \cite{ZM} gives the following result \cite{MD_neutrino}:

\begin{equation}
\Delta t_{GW \, , \, \mathrm{bounce}} \; \approx \; (0.1 \, \pm \, 0.4) \; \mathrm{ms}
\label{eq:GW_bounce}
\end{equation}
Therefore, from Eq. (\ref{eq:nu_bounce}) and (\ref{eq:GW_bounce}), the systematic difference between the neutrino signal and the GW peak is

\begin{equation}
\Delta t_{\nu_{\mathrm{e}} \, , \, \mathrm{GW}}^{\mathrm{sys}} \; \approx \; (3.4 \, \pm \, 0.7) \; \mathrm{ms}
\label{eq:nu_GW}
\end{equation}

This relation is valid for massless neutrinos propagating at the speed of light, like the GW signal. Conversely, if the neutrino mass $m_\nu$ is not zero, an additional delay appears which depends also on the neutrino energy $E_\nu$ and on the source distance $D$:

\begin{equation}
\Delta t_{\nu_{\mathrm{e}} \, , \, \mathrm{GW}}^{\mathrm{massive}} \; \approx \; 5.2 \, \mathrm{ms} \; \left( \frac{ D }{ 10 \, \mathrm{kpc} } \right) \; \left( \frac{ m_\nu \, c^2 }{ 1 \, \mathrm{eV} } \right)^2 \; \left( \frac{ 10 \, \mathrm{MeV} }{ E_\nu} \right)^2
\label{eq:nu_mass}
\end{equation}

Finally, a statistical error $\Delta t_{\nu_{\mathrm{e}} \, , \, \mathrm{GW}}^{\mathrm{stat}}$ is introduced by the two timing measurements and is given by the quadratic sum of the contributions defined in Eq. (\ref{eq:nu_stat}) and (\ref{eq:GW_stat}). Taking realistic values for the neutrino-GW coincidence detection, i.e. $N_e \approx 10$, a GW signal half-width $\omega=2$~ms triggered by the ALF filter with $\rho=5$ --, one gets $\Delta t_{\nu_{\mathrm{e}} \, , \, \mathrm{GW}}^{\mathrm{stat}}=1.1$~ms, which is the level required to be sensitive to the neutrino mass at the eV-scale.

\subsection{Core collapse chronology}

To briefly summarize the previous sections, one can try to establish a chronology for the different phases of the supernova explosion. First, two main signals are concentrated within a few ms during the bounce of the core: the GW burst itself and the electron neutrino short flash. Then, the thermal emissions of all neutrino flavors start and last for a few seconds. Finally, the optical emission is delayed by some hours.

\section{Detecting the GW signal}
\label{section:GW_detection}

Two main approaches for network data analysis can be considered \cite{arnaud_burst_3,PDB,Finn}. The first one looks for coincidences compatible in time between the different interferometers of the network. As no assumption on the source location in the sky is made, large coincidence windows must be opened for each pair of detectors, increasing the false alarm rate. On the other hand, a survey of the whole sky can be performed in one single process using this algorithm, called 'loose coincidence' in the following. Its performances -- studied in Ref. \cite{arnaud_burst_3} -- are briefly summarized in the next section \ref{subsection:direct_detection}. In addition to the excess of false alarms mentioned above, this method has another drawback: it does not use all the informations available, as detector outputs are only taken into account if they exceed a given threshold.

Due to these intrinsic limitations, the loose coincidence performances are limited. Therefore, other strategies must be investigated to improve the network potential and close supernovae are well-suitable to make these methods concrete. Indeed, these explosions provide additional emissions -- see Section \ref{subsection:nu_opt} --, more easily detected and bringing additional informations on the GW signal.

Consequently, in this favorable case, the interferometer outputs can be combined in a better way, and only reduced time series need to be analyzed. Section \ref{subsection:SN_confirmation} studies the potential of those 'confirmation' scenarios and focuses on three data analysis methods:

\begin{itemize}
\item 'Tight' coincidences, in which shorter coincidence windows are used as the delays between the different detectors can be accurately estimated with the knowledge of the source position in the sky.
\item The coherent approach\footnote{Another approach based on coherent analysis can be found in Ref. \cite{sylvestre_network}; it extends the power filter \cite{powermonit} in the 3-interferometer network made of Virgo and the two 4-km LIGO interferometers.}, merging all synchronized interferometer outputs, which optimally uses all the informations available in the network.
\item The 'OR' strategy -- at least one output over threshold in the network.
\end{itemize}

Appendix \ref{section:itf_network_simulation} gives more details on these different detection methods, such as on the simulations used to compute the efficiency curves presented in this article. Extensive informations can also be found in the original Ref. \cite{arnaud_burst_2,arnaud_burst_3}.
 
\subsection{Direct detection in coincidence}
\label{subsection:direct_detection}

Figure~4 summarizes the performances of the loose coincidence approach for a stand-alone GW search, both for the Virgo-LIGO network and for the full set of six interferometers. In all configurations, the optimal SNR is assumed to be $\rho_{\mathrm{max}}=10$, suited for a supernova at the Galactic center.

The main conclusion is that coincidences provide a high detection efficiency only at the expense of a high false alarm rate. Clearly, a network with only two (like the two LIGO interferometers, the best pair of detectors as they are well-aligned) or three instruments (Virgo-LIGO) is too small, and larger networks are required. With a six-component network, the situation is better even if only twofold and possibly threefold coincidences can trigger on real GW signals with an high confidence level. Therefore, a direct detection of a close supernova GW may be difficult if the signal has an optimal SNR around 10 or below. So alternative network strategies must be addressed. 

\subsection{Confirmation of GW emission}
\label{subsection:SN_confirmation}

This section presents strategies to look for the GW emission associated to a close type II supernova, previously located by neutrino and optical detectors. They assume that a suitable former analysis gave accurate estimations of the GW signal arrival time and of the source location in the sky. Using these informations allows one to cut more drastically on false alarms, while keeping the detection efficiency high.

In these scenarios, the trigger threshold needs to adjusted depending on the required goal of the data analysis, parameterized by two quantities:  the time window $\Delta T$ in which the search is performed, and the remaining false alarm probability $\tau$. As a supernova burst lasts only few ms, choosing $\Delta T = 10$~ms is accurate. Then, two different confidence levels can be considered: $\tau=10^{-6}$ -- corresponding to almost '$5\sigma$', a value needed to claim the discovery of a GW emission during a core collapse -- and $\tau=10^{-2}$, enough to simply study the coincidence between the GW and neutrino signals.

\subsubsection{Tight coincidences}

Figure~5 summarizes the performances of the tight coincidence strategies \cite{arnaud_burst_3} for $\rho_{\mathrm{max}}=10$. They are slightly better than in the loose coincidence case, as less false alarms pass the stronger compatibility test at a given threshold level. Yet, the efficiency gain remains limited. Configurations from two to six interferometers are studied. Table~4 compares the detection efficiencies of both loose -- see Figure~4 -- and tight coincidences at the false alarm probabilities of $\tau = 10^{-6}$ and $10^{-2}$ respectively.

\subsubsection{Network coherent analysis}

Figure~6 shows the efficiency curves for the coherent analysis in the two networks Virgo-LIGO and the set of six interferometers, assuming an optimal SNR $\rho_{\mathrm{max}}=10$. Note that the vertical scale is zero-suppressed. These curves are upper bounds of the network efficiency, as these algorithms use in an optimal way all the available information. Table~5 gives the detection efficiencies at the false alarm rate of $\tau = 10^{-6}$ and $\tau = 10^{-2}$: with six interferometers, the detection is almost certain, while it is already quite probable with the Virgo-LIGO network.

\subsubsection{OR network strategy}

As in the previous sections, Figure~7 focuses on two examples of networks: Virgo-LIGO and the full set of six interferometers. The 'OR' strategy' efficiency curves correspond to the least stringent condition: at least one detection in the network. Performances are a bit lower than those of the coherent analysis because of the large increase of the false alarm rate. Table~6 summarizes the detection efficiencies for $\tau = 10^{-6}$ and $\tau = 10^{-2}$.

\subsubsection{Summary}

The previous results show that despite the expected weakness of the supernova GW signals, a close event of this type has a significant probability to be confirmed in a network of interferometers: informations provided by the neutrino and the optical emissions strongly help the search analysis. A detection is very likely if the coherent approach is used, while the 'OR' strategy is also very powerful, although it would certainly be associated with a smaller confidence level as transient noises cannot be efficiently vetoed in this case. Even coincidences may give interesting detection efficiencies, especially for a network with six detectors.

\subsubsection{'OR' strategy and coherent analysis comparison}

As previously mentioned, the main advantage of the coherent analysis with respect to the 'OR' strategy appears when one takes into account non-stationnarities. A simple way to model them is to consider simulations in which all detector outputs except one are realizations of Gaussian and stationary noises, while the last one contains in addition a Gaussian peak of fixed optimal SNR (with $\rho_{\mathrm{max}}=10$ here), modeling the transient. Given the false alarm rate, one can estimate the number of fake detections $\epsilon_{\mathrm{fake}}$ associated with the noise burst which mimics a real GW signal. For the 'OR' strategy, $\epsilon_{\mathrm{fake}}$ is around 100\% from all false alarm rates until 1/week or even beyond, as no antenna pattern reduces the transient amplitude: all such events are detected with $\rho=\rho_{\mathrm{max}}$ in average! On the other hand, coherent analysis should discard most of them as the 'signal' is located in only one single detector while a real GW is in some sense global in the network.

To make this statement more concrete, one can compare the almost perfect fake efficiency of the 'OR' strategy with the two Tables~7 and~8. They present the values of $\epsilon_{\mathrm{fake}}$ in coherent analysis for different false alarm rates: Table~7 deals with the Virgo-LIGO network, and Table~8 with the full network. In both cases, the fake efficiency depends on the particular interferometer chosen to host the transient noise, as shown in these Tables. Indeed, $\epsilon_{\mathrm{fake}}$ is found to be smaller when the noise burst is located in a LIGO detector as transient events are in this case more easily vetoed thanks to the correlation between the two LIGO antenna patterns. Averaging over all interferometers, the false alarm rate of non-stationary signals is reduced by a factor 10-20 when operating coherently. Clearly, one can conclude from this comparison that the coherent analysis is much more efficient then the 'OR' strategy to distinguish transient noise from signal events, another advantage of this method.

\section{Exploiting the coincident detections of neutrino and GW emitted by a supernova}
\label{section:physics}

In addition to the great achievement that the detection of GW would represent, observing a supernova in coincidence in neutrino and GW detectors could lead to other interesting physical results. In this section, three possibilities are briefly considered: the determination of neutrino masses using the time-of-flight difference between GW and neutrino bursts, a test of the supernova collapse dynamics and the detection of  a black hole formation in case of an aborted supernova collapse.

\subsection{Neutrino masses}

As shown in Eq. (\ref{eq:nu_mass}), measuring the delay between the neutrino and GW bursts allows one to estimate the neutrino mass, assuming known the location of the progenitor of the supernova and that the GW propagate at the speed of light. Indeed, Ref. \cite{MD_neutrino} elaborates on the feasibility of this measurement. As the propagation of neutrinos can be affected by flavor oscillations -- see the discussion in Section~IV of Ref. \cite{MD_neutrino} for more details --, two cases for the $\nu_{\mathrm{e}}$ survival probability $P_{\mathrm{e}}$ are considered in this reference: 
\begin{itemize}
\item $P_{\mathrm{e}} = 0.5$, representative of situations where the $\nu_{\mathrm{e}}$'s survive;
\item $P_{\mathrm{e}} \sim 0$, when the $\nu_{\mathrm{e}}$ flux on Earth vanishes.
\end{itemize}

Four neutrino methods methods sensitive to timing have been studied: the first two look for $\nu_{\mathrm{e}}$ in SNO (1) and SuperKamiokande (2) respectively, while the two other use other flavors -- thermal $\bar{\nu}_{\mathrm{e}}$ in SuperKamiokande (3) or $\nu_{\mu,\tau}$ in SNO (4). In this way, even unfavorable scenarios in which all electron neutrinos oscillate to other flavors before reaching the detector are covered by methods (3) and (4). All methods use in addition the timing information provided by a GW interferometric detector.

Indeed, Figure~8 compares the sensitivities of the four methods mentioned above versus the supernova distance, given in kpc. All the curves end artificially at 13 kpc: at this distance, the probability to detect at least three neutrino events -- the minimum number required by methods (1) and (2) -- from the neutronization burst is still high (73\%) but it steadily decreases for more distant sources, only 55 (27)\% at 15 (17) kpc respectively. Methods (1) and (2) show sensitivities almost independent of the supernova distance and significally better than for methods (3) and (4). 

All sensitivities appear strongly degraded below a few kpc as the neutrino propagation delay scales like the supernova distance and the uncertainty on the delay becomes dominated by the systematic error from Eq. (\ref{eq:nu_GW}) -- see also Figure~9 which compares the different timing error contributions, assuming realistic values of $N_{\mathrm{e}}$ and $\rho$ for a supernova explosion at $D=10$~kpc. In contrast, the dashed line shows the sensitivity that the SNO detector alone can achieve using the charged-current reaction of method (1). For very close supernovae, its result is better than for methods (1) and (2), but it quickly degrades when the distance increases due to a lack of statistics for highly energetic neutrinos, needed to estimate the zero-mass timing. Therefore, the GW timing information is increasingly helpful when the supernova distance increases.

Finally, Figure~10 shows the neutrino mass sensitivity one can reach by combining these different methods. If the $\nu_e$'s survive $(P_e = 0.5)$, the four strategies can be combined while if they vanish $(P_e \sim 0)$, only the two last are useful.

Therefore, one can expect these analyzes to be sensitive at the 1-eV level or below, which would improve the current direct upper limit \cite{pdg2002} of 3~eV \cite{mainz,troitsk} on the $\nu_e$ mass. These new constraints would be of great interest as recent strongly convincing experimental results show that at least some neutrino flavors would have a nonzero mass -- see \cite{nu_mass_ICHEP02} for a recent synthesis on this topic. However, such supernova-based measurement would be direct and absolute, with a sensitivity almost independent of the source distance -- see both Figures~8 and~10 --, provided that both the neutrino and the GW signals are detected.

\subsection{Probing the supernova collapse}

\subsubsection{Collapse dynamics}

It may well be possible that, at the time of the next Galactic supernova, the neutrino masses are supposed to be better estimated: either a non-zero value is measured experimentally or, more likely, a more stringent upper limit is set, e.g. around 0.3~eV -- a target value for a future experiment measuring the end-point of the tritium $\beta$-decay spectrum \cite{KATRIN} which is also close to the indirect limit provided recently by the WMAP collaboration \cite{WMAP}: 0.23~eV at 95\% confidence limit. In that case, the small limit on the neutrino mass produces a negligible uncertainty on the relative timing between the GW and neutrino bursts.

For instance, let us consider a scenario in which $N_{\mathrm{e}}=10$ $\nu_e$ are detected from a supernova exploding at a distance $D=10$~kpc, while the GW counterpart, dominated by a millisecond peak, is seen with a SNR $\rho=5$ by a well-matched (Gaussian) template. Then, using Eq. (\ref{eq:nu_stat}) and (\ref{eq:GW_stat}), one gets numerically:

\begin{equation}
\Delta t_{\nu_{\mathrm{e}} \, , \, \mathrm{GW}}^{\mathrm{stat}} \; \approx \; 0.8 \; \mathrm{ms} \; \left( \frac{ D }{ 10 \; \mathrm{kpc} } \right) 
\end{equation}
a value comparable with the corresponding systematics -- see Eq. (\ref{eq:nu_GW}) and Figure~9, drawn assuming this scenario.

Therefore, assuming known the neutrino mass, the delay between GW and neutrino due to the supernova physics could be measured with a 20\% accuracy, providing unique information on the propagation time for the shock wave to reach the neutrinosphere. It could be compared with numerical simulation results to estimate either the propagation speed of the bounce or the neutrino mean free path, related to the core density, hence providing a valuable input to the state equation of nuclear matter.

This measurement would not be too much degraded if only an upper bound on the neutrino mass is known. Indeed, with $m_\nu \le 0.3$~eV, Eq. (\ref{eq:nu_mass}) becomes

\begin{equation}
\Delta t_{\nu_{\mathrm{e}} \, , \, \mathrm{GW}}^{\mathrm{massive}} \; \le \; 0.5 \, \mathrm{ms} \; \left( \frac{ D }{ 10 \, \mathrm{kpc} } \right) \; \left( \frac{ 10 \, \mathrm{MeV} }{ E_\nu} \right)^2
\end{equation}
This measurement bias is smaller than the statistical timing error computed above.

\subsubsection{Supernova collapse to a black hole}

Another interesting possibility would be an inner core collapse followed by a phase of matter accretion from the outer mantle \cite{beacom}. In this case, the core mass grows and finally collapses into a black hole in a timescale of about 0.5~s. The signature of this black hole formation would be a fast -- $\sim 0.5$~ms -- cutoff in the neutrino signal, allowing a precise timing of the event.

The excited black hole is expected to return to equilibrium by emitting GW through its normal modes of oscillation \cite{BH_normal_modes}. These signals are damped oscillations which could be efficiently tracked with matched filtering \cite{creighton,arnaud_tiling,tama_DS}. In addition, the precise timing location provided by the neutrino signal would limit the length of the data to be analyzed, thus allowing to use low thresholds with high confidence levels, and efficiencies similar to those quoted in Section \ref{subsection:SN_confirmation} should be reachable as the corresponding SNR are high enough to expect detections in the Local Group \cite{SP,FP}. Detecting the abrupt neutrino disappearance followed by the characteristic high-frequency ring-down GW signal would be a clear and direct proof of a black hole formation.

\section{Conclusion}

In the next few years, a network of giant interferometric detectors will try to detect GW signals for the first time. Various sources are theoretically expected, among which the Type II supernovae. Detecting at least one GW emission associated to such explosion would bring unique information, extending well beyond the proof of the GW existence. Indeed, it would allow to measure or set bounds on neutrino masses, to improve the knowledge on the supernova physics by precise timing of the collapse events, or even to provide the clear signature of a black hole formation.

The most difficult aspect of such analyzes is clearly the GW itself: other emissions (neutrino and optical) will be easier and less ambiguous to detect. As a very small fraction of the core mass is converted into GW, the signal amplitude is low and only close events should be visible in interferometers. Consequently, these events are rare and at most a few of them are expected to occur during the lifetime of GW detectors. Therefore, finding methods as efficient as possible to detect the GW waveform is in our view a major goal of the data analysis preparation.

This article compares three possible strategies (the coincidence method, the coherent approach and the 'OR' network strategy), thanks to numerical simulations of network filtering. The network modeling aims at being realistic, although some hypotheses have been made to simplify the computation. The main results are twofold: on the one-hand, a direct coincidence detection involving only interferometers does not appear very likely, even in a network of six instruments. But, on the other hand, taking advantage of the other emissions associated with the supernova explosion strongly increases the network performances, making likely the confirmation of the core collapse in the GW sector, hence potentially providing an unambiguous discovery of GW. Efficiencies above 70-80\% are possible in most of the configurations and in favourable situations they may even reach almost 100\%.

The work presented here aims at providing some directions which should be explored by further detailed studies. In particular, defining and testing in real conditions the data analysis procedures which have been sketched in this article will be a major collaborative challenge to the current GW detector community in the future. 

\appendix

\section{Appendix: interferometer network simulation}
\label{section:itf_network_simulation}

This appendix summarizes the basic definitions and the main assumptions used in the core of the article about GW detection. In particular, the hypothesis corresponding to the network data analysis model are given.

\subsection{Filtering methods}

The main difficulty of the GW data analysis is to be able to detect GW signals occurring at random time in a noisy background. Therefore on-line filters will be used to select -- at a manageable rate, at most few events per hour and per filtering method -- segments of data potentially including a real waveform. This reduced set of data will then be re-analyzed off-line in more details, in order to tag it as GW event or noise fluctuation.

Whatever the analysis considered, its first step is always to see whether or not a particular quantity, called signal-to-noise ratio (SNR in short) exceeds a threshold, previously tuned to a given selection rate. This SNR is generically defined by the following equation:

\begin{equation}
\mathrm{SNR}(t) \; = \; \frac{ \mathrm{Filter \, output \, at \, time \, t} }{ \mathrm{RMS \, of \, the \, filter \, output \, with \, noise \, only} }
\end{equation}

Among all filters, the best one is the Wiener (or matched) filter: it gives the highest SNR for a given signal \cite{WZ} and has in addition the lowest false dismissal rate for a given threshold \cite{MF_optimal}. However, using it requires an accurate knowledge of the searched signal: as soon as the real signal and the template -- i.e. the specific filter waveform with which the detector output is correlated -- do not match exactly, the SNR decreases dramatically. 

Due to this lack of robustness, handling an efficient matched filtering is usually computationally expensive. Indeed, even if the signal shape is well-known, the true waveforms still depends on a vector of parameters (e.g. the mass, the angular momentum or the source location in the sky) specific to the source -- and thus unknown. So, the whole set of values physically allowed for these quantities (called the parameter space) must be accurately covered not to loose the real signal. A given template corresponding to a particular vector of the parameter space is efficient in only a small area and so many different ones must be used in parallel. A good tiling of the parameter space must have no hole and a reduced number of filters to limit its CPU cost. Computing such sets of templates is a difficult problem which has already been studied in the literature; for a presentation of the framework used in these studies, one can see Ref. \cite{Owen}.

As an example, the coherent analysis presented in this article can be seen as a matched filter search with three parameters: the signal half-width $\omega$ and the two coordinates locating the GW source on the celestial sphere. This method is also used to search signals emitted during the in-spiral phase of coalescing compact binaries (neutron stars or black holes), see e.g. \cite{Blanchet}. On the other hand, burst waveforms are most of the time poorly known. Therefore, alternative methods have been designed -- see e.g. \cite{pradier,powermonit,arnaud_burst_1,bala,mohanty,vicere,sylvestre} --, all aiming at being robust and efficient over the widest possible class of signals.

\subsection{GW signal}

After interaction of a GW with an interferometric detector, the signal $h(t)$ is a linear combination of its two polarizations $h_+(t)$ and $h_\times(t)$.

\begin{equation}
h(t) \; = \; F_+(t) \, h_+(t) \; + \; F_\times(t) \, h_\times(t)
\label{eq:response}
\end{equation}

The beam pattern functions $F_+$ and $F_\times$ describe the spatial detector response, which depends on the relative position of the source in the sky with respect to the instrument \cite{thorne87}. They are both below 1 in absolute value, which shows that the GW signal is reduced in general by this non-uniform sensitivity. In addition, the limited knowledge in the waveform (especially for what concern GW bursts) creates additional losses in SNR during the detection process.

Nevertheless, one needs to quantify the GW strength in a way independent from the interferometer location, or from the data analysis algorithm used to search it. So, one introduces the optimal SNR $\rho_{\mathrm{max}}$ \cite{arnaud_burst_2} defined as the mean output of the Wiener filtering (optimal among all data analysis methods as the filter is matched to the searched signal) in the presence of signal, and for a source optimally orientated.

As GW bursts are poorly-modeled, a choice of the waveform must be done in order to make the simulations concrete. Like in previous articles dealing with the same topic \cite{arnaud_burst_2,arnaud_burst_3}, one assumes here that the signal is a Gaussian peak of half-width $\omega$. In addition to this parameter, it is completely defined by its maximal amplitude $h_{\mathrm{peak}}$ scaled at a given reference distance, $D=1$~Mpc for instance. Thus, one has:

\begin{equation}
h(t) \; = \; \left( \frac{ \mathrm{ 1 Mpc } }{ D } \right) \; h_{\mathrm{peak}} \; \exp \left( - \frac{ t^2 }{ 2 \, \omega^2 } \right)
\end{equation}
$h_{\mathrm{peak}}$ and the optimal SNR $\rho_{\mathrm{max}}$ are connected by Eq. (\ref{eq:rho_max}) for the white noise case.

A multiplicative reduction factor $0<\kappa_{\mathrm{eff}} \lesssim 1$ can be added to reduce the amplitude of $h(t)$; this takes into account the fact that this particular choice of the burst waveform is only partially reproducing the real GW signal. 

To search the GW signal, the matched filtering procedure is used as Ref. \cite{arnaud_burst_1,arnaud_these} show that a precise coverage of a large range for the parameter $\omega$ can be achieved with a very small number of templates. Given the network configuration, the performances of a particular filter applied to some GW signal are nicely summarized on a single plot, the efficiency curve, which shows the detection efficiency as a function of the normalized false alarm rate $\tau_{\mathrm{norm}}$. The unit of the normalized false alarm rate is 'per bin'; for instance, assuming like in this article a sampling frequency $f_{\mathrm{samp}}=20$~kHz, $\tau_{\mathrm{norm}} = 1.39 \times 10^{-8}$ corresponds to 1 false alarm/hour in average.

\subsection{Interferometer network}

The non-uniformity of the beam pattern functions also implies that a given GW signal will be seen differently in amplitude and shape in distant detectors. This direct consequence of the non-optimal interferometer pattern is the main limitation of all network data analysis methods.

Such algorithms belong to two generic categories: coincidence or coherent filtering. While the former looks for compatibility in sets of data separately triggered by the different components of the interferometer network, the latter uses a prior hypothesis on the source location in the sky to combine all the data flows in a single one, simultaneously analyzed. Coincidences pay the price of the binary treatment of events (a GW is either present or absent), and are thus less efficient than coherent analyzes, which also benefit from signal contributions below the single detector threshold level \cite{arnaud_burst_3}.

In addition, a less sensitive interferometer will loose a larger fraction of GW signals, even in case of optimal orientation: it will only weakly contribute to the network potential. Therefore, this makes the relative weights of detectors crucial for the network design. Here, we keep the hypothesis of Ref. \cite{arnaud_burst_2,arnaud_burst_3} and assume that all instruments are identical. This choice has two main motivations: first, the future experimental sensitivities of real interferometers are still unknown, which makes difficult to define a realistic and precise hierarchy between them. Moreover, this uniformity of interferometers can be seen as a mid-term goal of the worldwide GW community, as an efficient network should only include detectors with near sensitivities.

\subsection{Simulation main characteristics}

So, networks are compared through the geographical complementarity of their components. All detectors are assumed to be located where the first generation of interferometers are currently being operated or planned. The response of any instrument to a given GW signal depends on four angles: its latitude $l$, its longitude $L$ -- positive westward by convention --, its arm separation $\chi$ and finally its local orientation $\gamma$, e.g. with respect to the local North-South direction. The ACIGA orientation, not yet decided, has been optimized in order to maximize the performances of the full network of six interferometers \cite{arnaud_burst_2,Searle}. All detector data used as simulation inputs are summarized in Table~9.

Like in Ref. \cite{arnaud_burst_3}, two networks are extensively used as examples of interferometer configurations: the three-detector network Virgo plus the two 4-km LIGO instruments and the full network of the six interferometers (ACIGA, GEO600, the two 4-km LIGO instruments, TAMA300 and Virgo).

All averaged GW detection efficiencies are computed in simulation, assuming a uniform distribution of sources over the sky. Indeed, each source is defined by three variables: its two celestial sphere coordinates -- the right ascension $\alpha$ and the declination $\sin\delta$ -- and the polarization angle $\psi$, drawn from uniform distributions in $[-\pi;\pi]$, $[-1;1]$ and $[-\pi;\pi]$ respectively.

The simulation of the network data analysis methods -- coincidence and coherent filtering -- is identical to the procedures described in Ref. \cite{arnaud_burst_2,arnaud_burst_3} -- to which the reader is referred for further details.

Some additional information is summarized below:
\begin{itemize}
\item Interferometer noises are assumed to be Gaussian, white and uncorrelated with the same RMS. This is of course an imperfect model of real data but whitening procedures \cite{cuoco1,cuoco2}, removal line algorithms \cite{searle2} and veto algorithms \cite{ligo_S1} aiming at removing non-stationary datasets should make the interferometer outputs close to this ideal case. In addition, the power spectrum density of interferometer noises is quite flat in the expected supernova GW frequency range centered in the kHz range.
\item Noise and signal are sampled at $f_{\mathrm{samp}}=20$~kHz and the signal arrival time is chosen at random in the analysis window -- i.e. it does not match any sampling time. Delays due to the finite GW speed are also taken into account when a GW signal is generated in a network of detectors.
\item A data set in interferometer $I^0$ is said to have triggered if at least one of the filter output bins exceeds a threshold $\eta$, corresponding to a chosen false alarm rate $\tau$. As consecutive filter outputs are strongly correlated, only the maximum bin above the threshold is taken into account. Its value $\rho^0$ and its occurrence time $t^0$ define an event. In addition, the $\rho^0$ value allows one to estimate the error $\Delta t^0_{\mathrm{RMS}}$ on the GW timing, due to noise fluctuations \cite{arnaud_burst_2,arnaud_burst_3}: the higher $\rho^0$, the better the GW localization.
\item All coincidence procedures test the compatibility between pairs of events, located in interferometers $I^i$ and $I^j$. One introduces the delay between the events $\Delta t^{ij} = t^j - t^i$ and its error $\Delta t^{ij}_{\mathrm{RMS}}$, computed by summing in quadrature $\Delta t^i_{\mathrm{RMS}}$ and $\Delta t^j_{\mathrm{RMS}}$. For loose coincidences -- when the source location is unknown --, one requires
\begin{equation}
\nonumber
\left| \, \Delta t^{ij} \, \right| \; \le \; 1 \; \times \; \Delta t^{ij}_{\mathrm{max}} \; + \; \Delta t^{ij}_{\mathrm{RMS}}
\end{equation}
to have compatible events, $\Delta t^{ij}_{\mathrm{max}}$ being the maximum delay between the two detectors, reached when the GW arrives along the direction $\overrightarrow{I^iI^j}$. On the other hand, with an estimation of the source location -- enabling tight coincidences to be required --, the delay between events $\Delta t^{ij}_{\mathrm{est}}$ can be estimated a priori. Therefore, the compatibility condition becomes:
\begin{equation}
\nonumber
\left| \, \Delta t^{ij} \; - \; \Delta t^{ij}_{\mathrm{est}} \, \right| \; \le \; 3 \; \times \; \Delta t^{ij}_{\mathrm{RMS}}
\end{equation}
In the two previous equations, the coefficients 1 and 3 in front of the delay error $\Delta t^{ij}_{\mathrm{RMS}}$ have been tuned in simulation to reach the best compromise between low false alarm rates and high detection efficiencies.
\item For coherent analysis, the source location is assumed to be known which allows one to synchronize in time the interferometer data segments.
\item Finally, for the 'OR' strategy the detection condition is the weakest possible: one requires only one detection among all the network components.
\end{itemize}




\newpage

\begin{table}[here!]
\begin{center}
\begin{tabular}{|c|c|c|c|}\hline
 Galaxy Type               & Ia          & Ib/c        & II \\ \hline
 E-S0 (Elliptic-Lenticular)& $0.18 \pm 0.06$ &  $< 0.01$ &  $< 0.02$ \\ \hline
 S0a-Sb (Spiral)           & $0.18 \pm 0.07$ & $0.11 \pm 0.06$ & $0.42 \pm 0.19$\\ \hline
 Sbc-Sd (Barred Spiral)    & $0.21 \pm 0.08$ & $0.14 \pm 0.07$ & $0.86 \pm 0.35$ \\ \hline
 Others (Irregular, Dwarfs)& $0.40 \pm 0.16$ & $0.22 \pm 0.16$ & $0.65 \pm 0.39$ \\ \hline
 All             & $0.20 \pm 0.06$ & $0.08 \pm 0.04$ & $0.40 \pm 0.19$ \\ \hline
\end{tabular}
\vskip 0.2truecm
Table~1: Supernova rates (SNu) for different supernova and galactic types \cite{Cappellaro}.
\end{center}
\end{table}

\vskip 1truecm

\begin{table}[here!]
\begin{center}
\begin{tabular}{|c|c|c|c|}
\hline Detector & SuperKamiokande & SNO & LVD \\
\hline $\nu_e$ & 91 & 132 & 3 \\
\hline $\bar{\nu}_e$ & 4300 & 442 & 135 \\
\hline $\nu_\mu$, $\nu_\tau$ & (40) & 207 & (7) \\
\hline $\nu_e$ flash & 12 & 9 & 0.4 \\
\hline All & 4430 & 781 & 146 \\
\hline
\end{tabular}
\vskip 0.2truecm
Table~2: Neutrino event rate for a supernova at 10 kpc. The numbers in parentheses correspond to interacting neutrinos, but without flavor identification.
\end{center}
\end{table}

\vskip 1truecm

\begin{table}[here!]
\begin{center}
\begin{tabular}{|c|c|c|c|}
\hline Filter & Wiener filter & Mean filter & ALF \\ 
\hline $K$ & 14.5\% & 24.6\% & 25.3\% \\
\hline $\alpha$ & 1 & 0.68 & 0.71 \\
\hline
\end{tabular}
\vskip 0.2truecm
Table~3: GW filter timing parameters.
\end{center}
\end{table}

\vskip 1truecm

\begin{table}[here!]
\begin{center}
\begin{tabular}{|c|c|c|c|c|c|c|}
\hline Configuration & LIGO & Virgo-LIGO & Virgo-LIGO & Full network & Full network \\
& network & 2/3 & 3/3 & 2/6 & 3/6 \\
\hline Loose performances & 28\% & 36\% & 15\% & 73\% & 56\% \\
\hline Tight performances & 31\% & 40\% & 17\% & 81\% & 64\% \\
\hline
\end{tabular}
\vskip 0.2truecm
False alarm probability $\tau = 10^{-6}$ \\
\vskip 0.2truecm
\begin{tabular}{|c|c|c|c|c|c|c|}
\hline Configuration & LIGO & Virgo-LIGO & Virgo-LIGO & Full network & Full network \\
& network & 2/3 & 3/3 & 2/6 & 3/6 \\
\hline Loose performances & 50\% & 65\% & 34\% & 96\% & 85\% \\
\hline Tight performances & 52\% & 67\% & 35\% & 97\% & 88\% \\
\hline
\end{tabular}
\vskip 0.2truecm
False alarm probability $\tau = 10^{-2}$
\vskip 0.2truecm
Table~4: Loose and tight coincidence confirmation performances for two false alarm probabilities.
\end{center}
\end{table}

\vskip 1truecm

\begin{table}[here!]
\begin{center}
\begin{tabular}{|c|c|c|}
\hline Network & Optimal SNR $\rho_{\mathrm{max}}$ & Detection efficiency \\
\hline & 5 & 8\% \\
Virgo-LIGO & 7.5 & 41\%\\
& 10 & 65\% \\
\hline & 5 & 33\% \\
Six-interferometer network  & 7.5 & 90\% \\
& 10 & 98.5\% \\
\hline
\end{tabular}
\vskip 0.2truecm
False alarm probability $\tau = 10^{-6}$ \\
\vskip 0.2truecm
\begin{tabular}{|c|c|c|}
\hline Network & Optimal SNR $\rho_{\mathrm{max}}$ & Detection efficiency \\
\hline & 5 & 56\% \\
Virgo-LIGO & 7.5 & 80\%\\
& 10 & 90\% \\
\hline & 5 & 93\% \\
Six-interferometer network  & 7.5 & 99\% \\
& 10 & 100\% \\
\hline
\end{tabular}
\vskip 0.2truecm
False alarm probability $\tau = 10^{-2}$
\vskip 0.2truecm
Table~5: Coherent analysis confirmation efficiencies.
\end{center}
\end{table}

\vskip 1truecm

\begin{table}[here!]
\begin{center}
\begin{tabular}{|c|c|c|}
\hline Network & Optimal SNR $\rho_{\mathrm{max}}$ & Detection efficiency \\
\hline & 5 & 2\% \\
Virgo-LIGO & 7.5 & 21\%\\
& 10 & 47\% \\
\hline & 5 & 3\% \\
Six-interferometer network  & 7.5 & 39\% \\
& 10 & 82\% \\
\hline
\end{tabular}
\vskip 0.2truecm
False alarm probability $\tau = 10^{-6}$ \\
\vskip 0.2truecm
\begin{tabular}{|c|c|c|}
\hline Network & Optimal SNR $\rho_{\mathrm{max}}$ & Detection efficiency \\
\hline & 5 & 39\% \\
Virgo-LIGO & 7.5 & 70\%\\
& 10 & 84\% \\
\hline & 5 & 63\% \\
Six-interferometer network  & 7.5 & 96\% \\
& 10 & 99\% \\
\hline
\end{tabular}
\vskip 0.2truecm
False alarm probability $\tau = 10^{-2}$
\vskip 0.2truecm
Table~6: 'OR' strategy confirmation efficiencies.
\end{center} 
\end{table}

\begin{table}[here!]
\begin{center}
\begin{tabular}{|c|c|c|c|}
\hline Transient location & Virgo & LIGO Hanford & LIGO Livingston \\
\hline $\tau_{\mathrm{norm}}=1/$week & 14.5\% & 7.0\% & 8.3\% \\
\hline $\tau_{\mathrm{norm}}=1/$day & 17.9\% & 9.8\% & 11.0\% \\
\hline $\tau_{\mathrm{norm}}=1/$hour & 22.7\% & 15.2\% & 16.1\% \\
\hline
\end{tabular}
\vskip 0.2truecm
Table~7: Transient fake detections for the coherent analysis in the Virgo-LIGO network.
\end{center}
\end{table}

\vskip 1truecm

\begin{table}[here!]
\begin{center}
\begin{tabular}{|c|c|c|c|c|c|c|}
\hline Transient & Virgo & LIGO & LIGO & GEO600 & TAMA300 & ACIGA \\
location & & Hanford & Livingston & & & \\
\hline $\tau_{\mathrm{norm}}=1/$week & 6.5\% & 2.7\% & 3.6\% & 6.6\% & 7.8\% & 6.8\% \\
\hline $\tau_{\mathrm{norm}}=1/$day & 8.4\% & 3.8\% & 4.8\% & 8.2\% & 8.6\% & 9.1\% \\
\hline $\tau_{\mathrm{norm}}=1/$hour & 10.7\% & 5.7\% & 7.6\% & 10.7\% & 11.5\% & 12.2\% \\
\hline
\end{tabular}
\vskip 0.2truecm
Table~8: Transient fake detections for the coherent analysis in the six-interferometer network.
\end{center}
\end{table}

\begin{table}[here!]
\begin{center}
\begin{tabular}{|c|c|c|c|c|}
\hline Detector & Latitude $l$ & Longitude $L$ & Arms 'separation' $\chi$ & Local orientation $\gamma$ \\
\hline ACIGA & -31.4 & -115.7 & 90.0 & 0.0 (optimized) \\
\hline GEO600 & 52.3 & -9.8 & 94.3 & 158.8 \\
\hline LIGO Hanford & 46.5 & 119.4 & 90.0 & 261.8 \\
\hline LIGO Livingston & 30.6 & 90.8 & 90.0 & 333.0 \\ 
\hline TAMA300 & 35.7 & -139.5 & 90.0 & 315.0 \\
\hline VIRGO & 43.6 & -10.5 & 90.0 & 206.5 \\
\hline
\end{tabular}
\vskip 0.2truecm
Table~9: First generation interferometer characteristics; all angles are given in degrees.
\end{center}
\end{table}

\newpage

\begin{figure}[here!]
\centerline{\epsfig{file=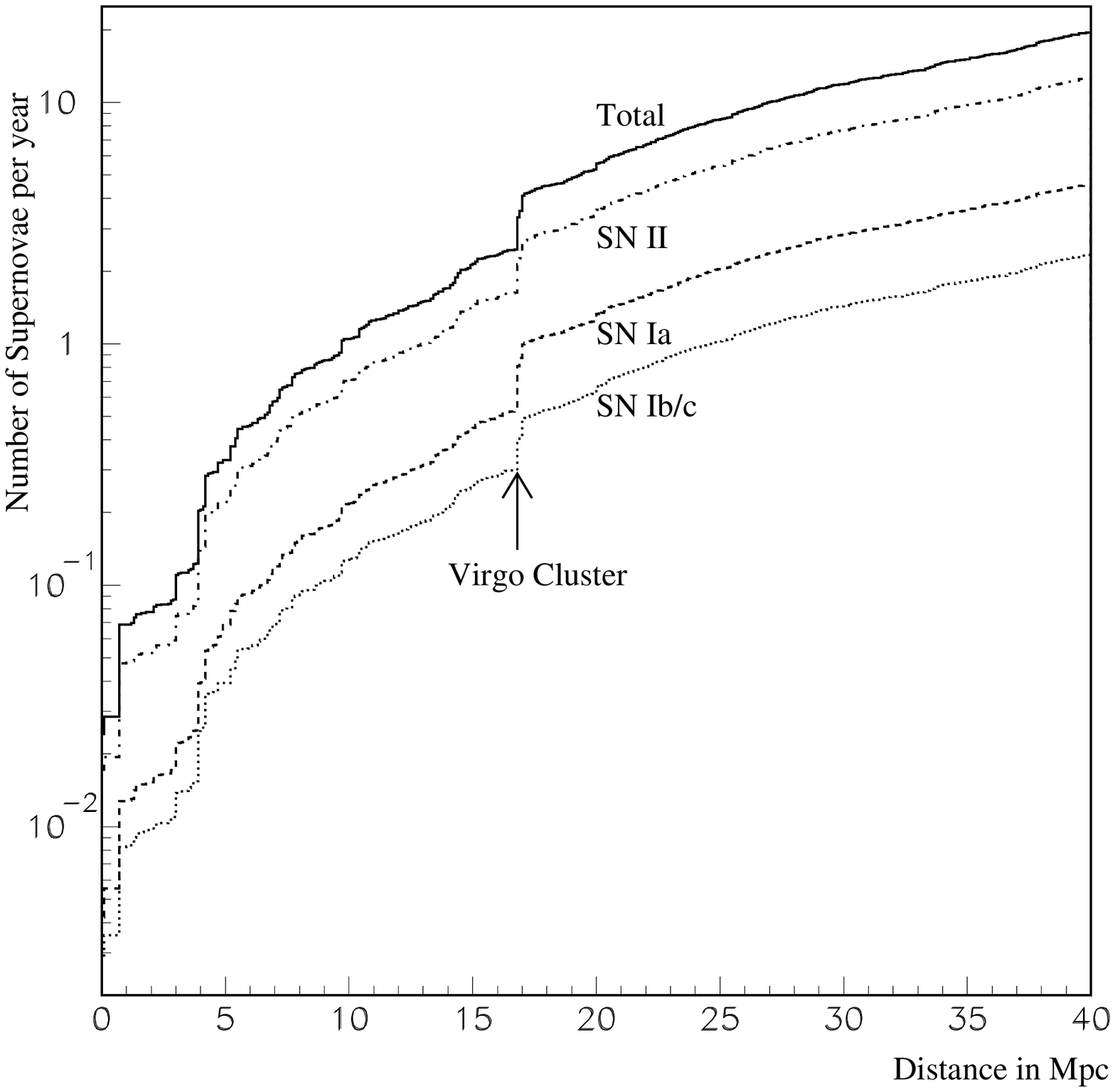,width=20cm}} 
Figure 1: Number of supernova per year as function of the survey distance in Mpc, estimated from the Tully catalog \cite{Tully}. The assumed Hubble constant value is $H_0=75$ km/sec/Mpc.
\end{figure}

\begin{figure}[here!]
\centerline{\epsfig{file=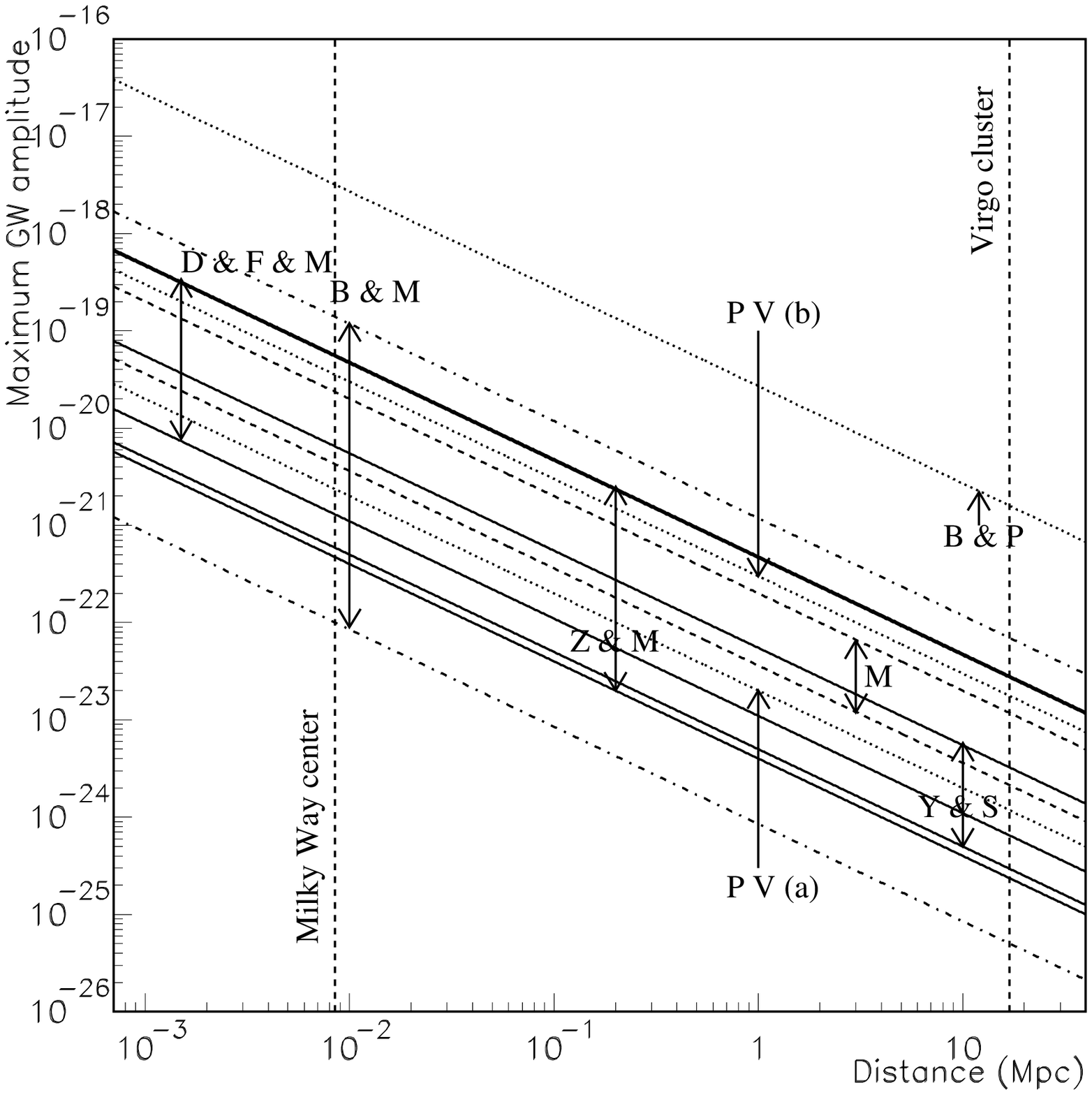,width=20cm}}
Figure~2: Comparison of the maximal amplitudes predicted for supernova events versus the source distance; the GW scale like 1 / distance. Six curves correspond to simulation results and are labeled by the initials of their authors: {\bf B \& M} \cite{BM}, {\bf Z \& M} \cite{ZM}, {\bf D \& F \& M} \cite{DFM}, {\bf M} \cite{GW_simul_1}, {\bf B \& P} \cite{GW_simul_3} and {\bf Y \& S} \cite{GW_simul_4}. The two remaining ones are limits provided by analysis of measured pulsar velocities: {\bf PV (a)} \cite{pulsar_velocity_1} and {\bf PV (b)} \cite{pulsar_velocity_2}.
\end{figure}

\begin{figure}[here!]
\centerline{\epsfig{file=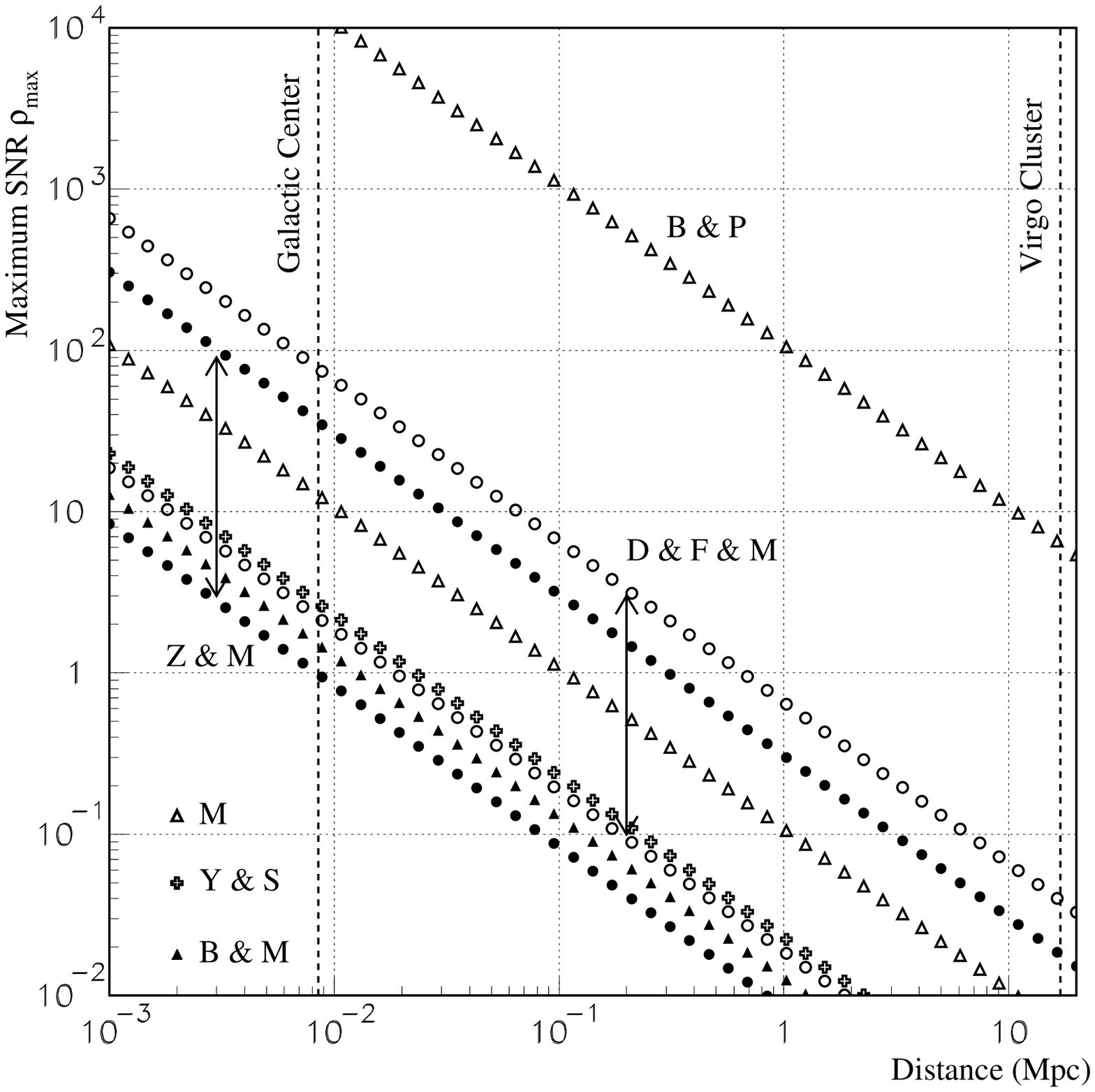,width=20cm}}
Figure~3: Evolution of the optimal SNR $\rho_{\mathrm{max}}$ versus the source distance $D$ for the different simulation results. The plots are based on the model developed in Section \ref{subsubsection:E_comp} and the values of the parameters correspond to those given in Eq. (\ref{eq:rho_vs_E}). For two numerical studies -- {\bf Z \& M} and {\bf D \& F \& M} --, the curves corresponding to the extreme simulated $E_{\mathrm{GW}}$ are drawn while only one representative value of the emitted GW energy is presented for the other simulations -- {\bf B \& P}, {\bf M}, {\bf Y \& S} and {\bf B \& M}.
\end{figure}

\begin{figure}[here!]
\centerline{\epsfig{file=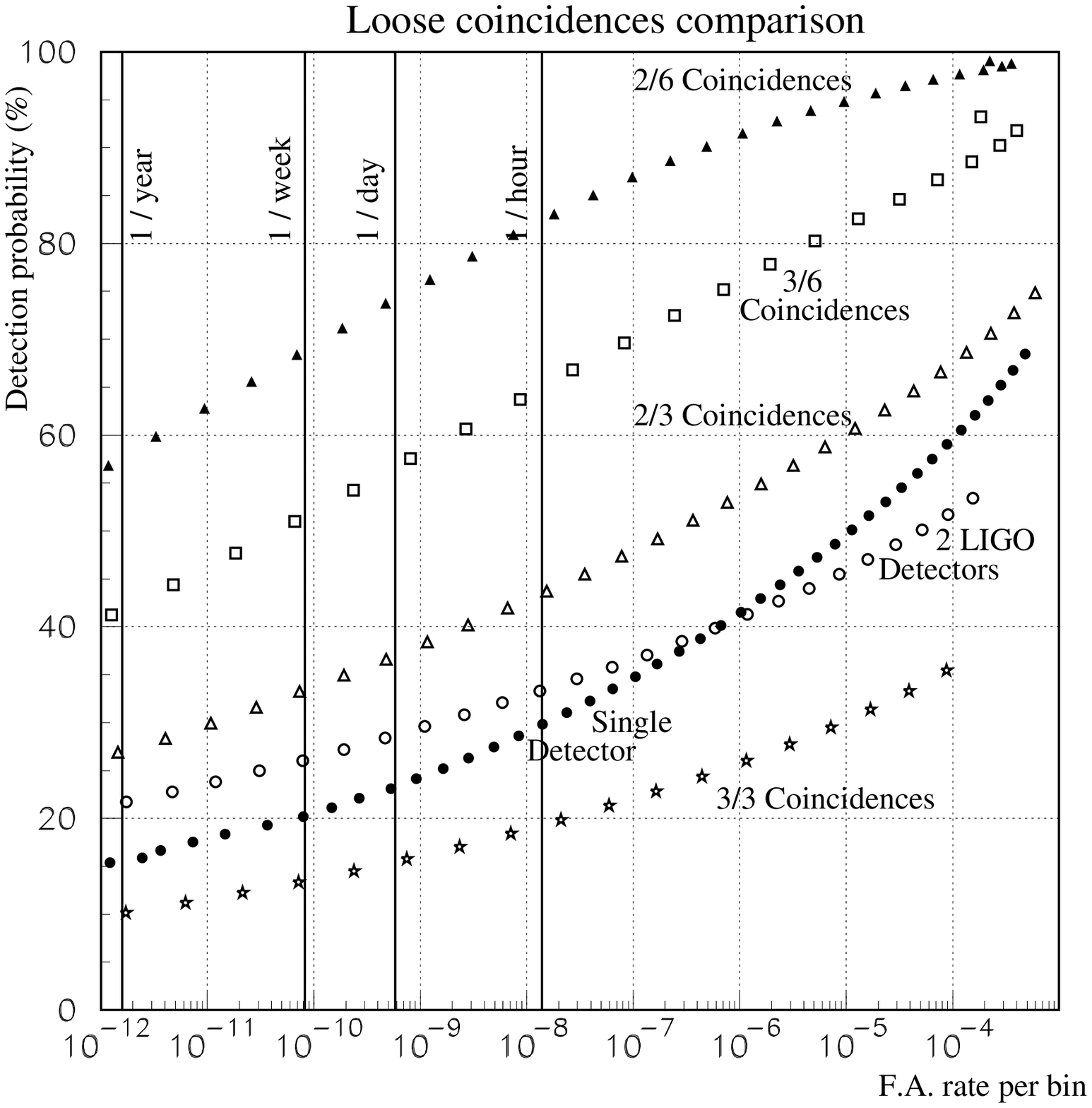,width=20cm}}
Figure~4: Comparison of efficiency curves ($\rho_{\mathrm{max}}=10$) corresponding to various coincidence strategies: single detector case, LIGO coincidences, twofold and threefold detections in the Virgo-LIGO network and in the full network of six interferometers.
\end{figure}

\begin{figure}[here!]
\centerline{\epsfig{file=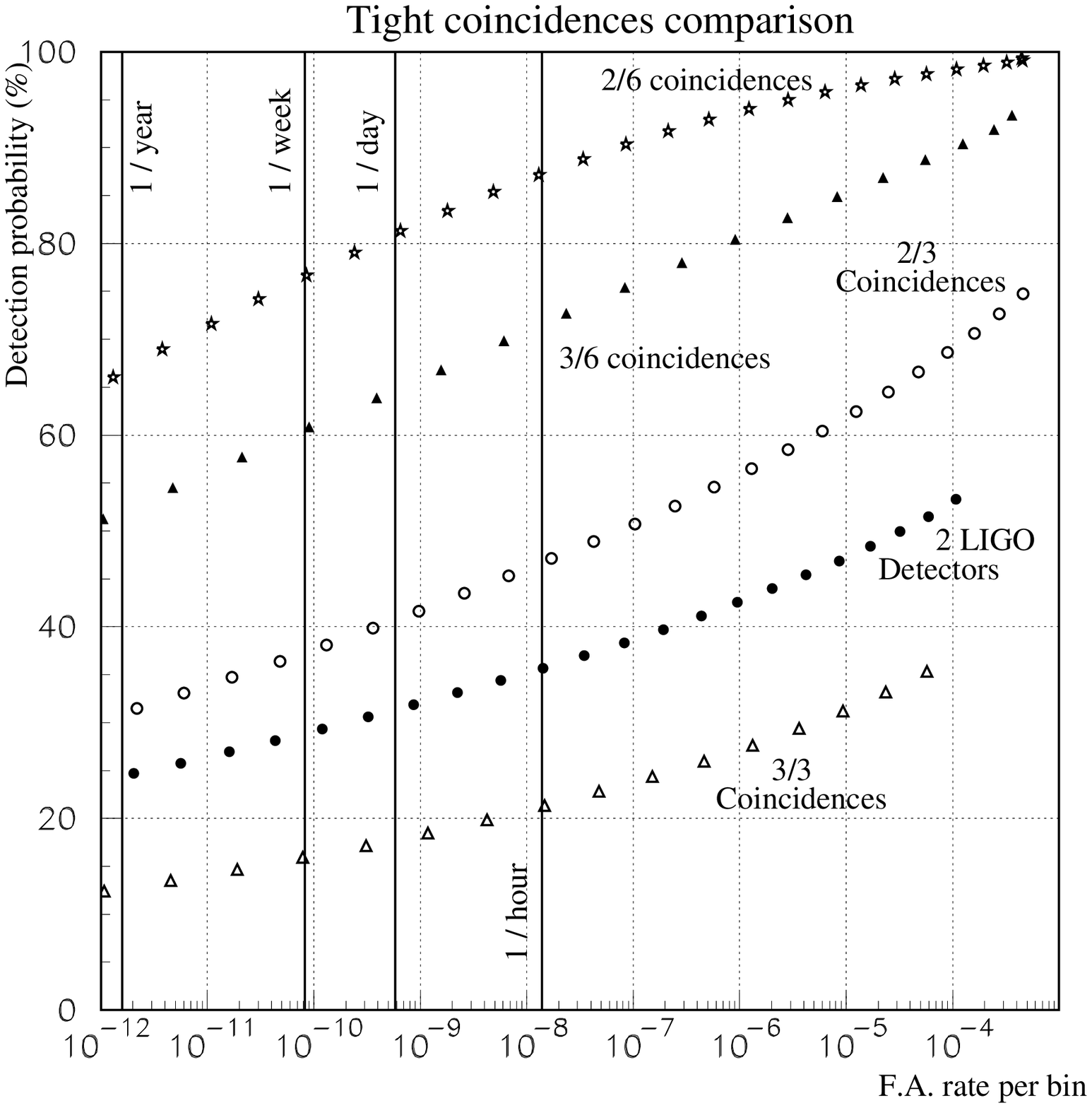,width=20cm}}
Figure~5: Comparison of efficiency curves ($\rho_{\mathrm{max}}=10$) corresponding to various tight coincidence strategies: LIGO coincidences, twofold and threefold detections in the Virgo-LIGO network and in the full network of six interferometers.
\end{figure}

\begin{figure}[here!]
\centerline{\epsfig{file=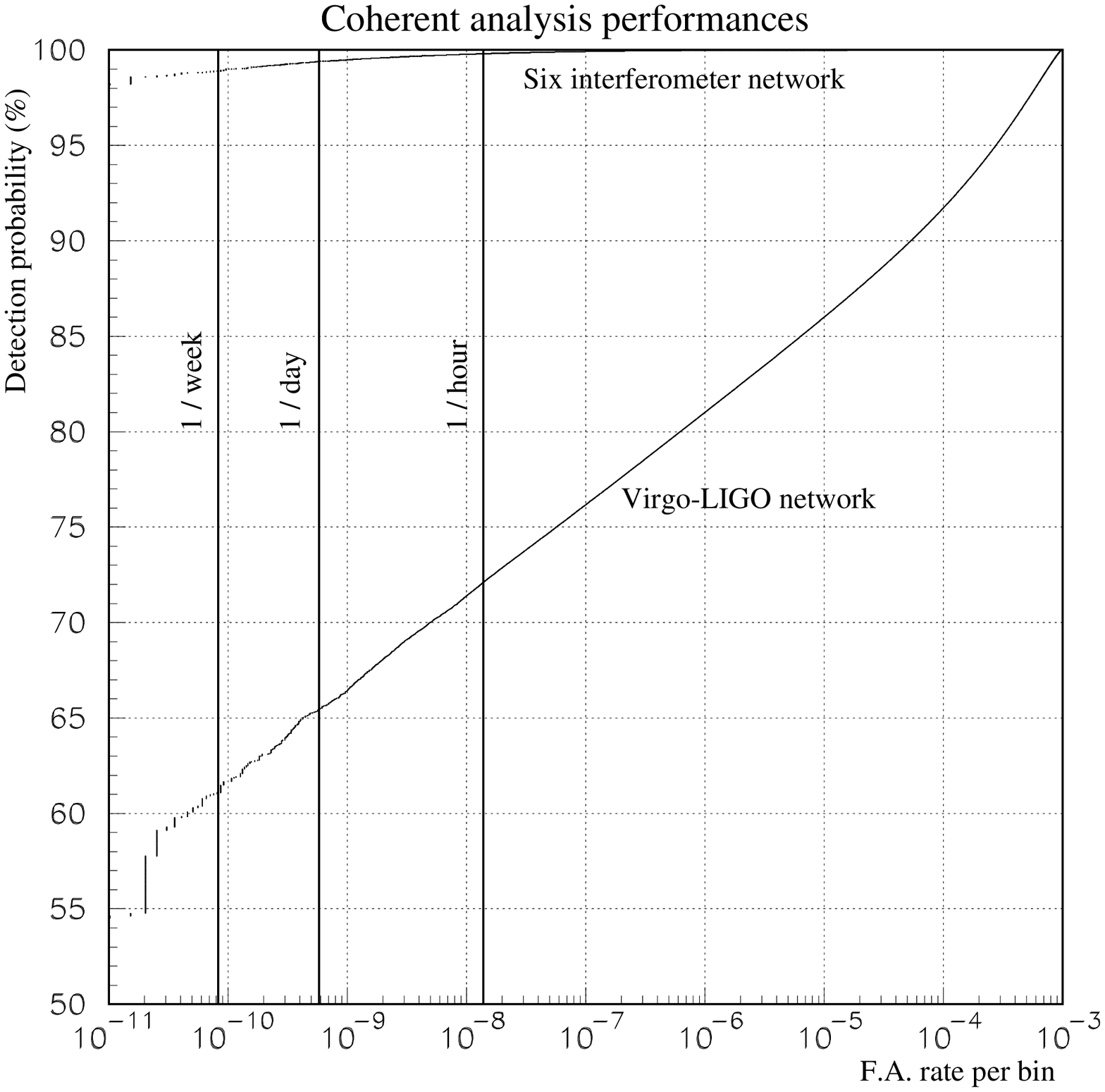,width=20cm}}
Figure~6: Coherent analysis performances for the Virgo-LIGO network and the full set of six interferometers for a GW signal with $\rho_{\mathrm{max}}=10$. Note the zero-suppressed vertical scale of the plot.
\end{figure}

\begin{figure}[here!]
\centerline{\epsfig{file=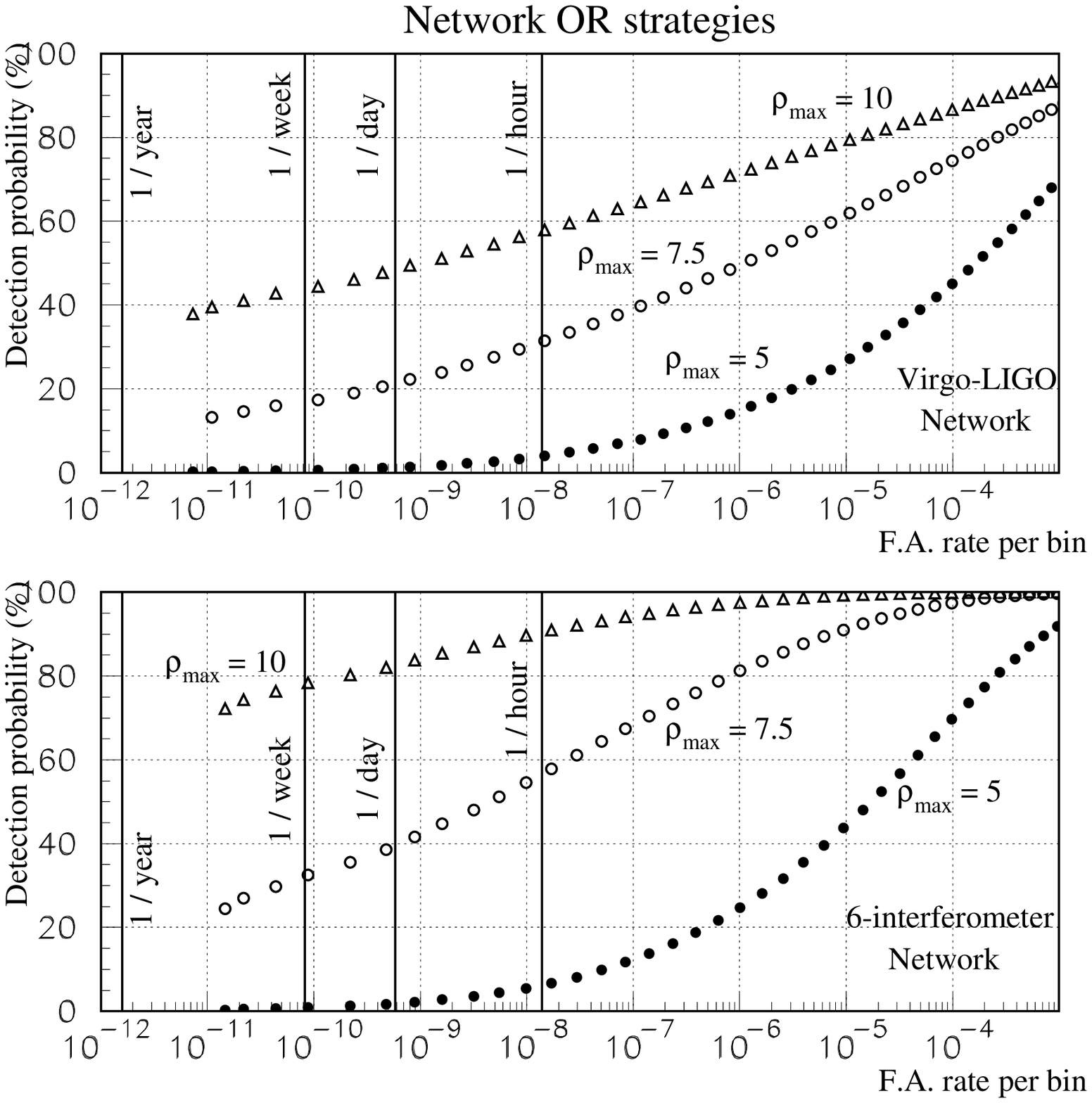,width=20cm}}
Figure~7: 'OR' strategy efficiency curves for the Virgo-LIGO network and for the full set of interferometers. In each plot, three values of the optimal SNR $\rho_{\mathrm{max}}$ are considered: 5, 7.5 and 10.
\end{figure}

\begin{figure}[here!]
\centerline{\epsfig{file=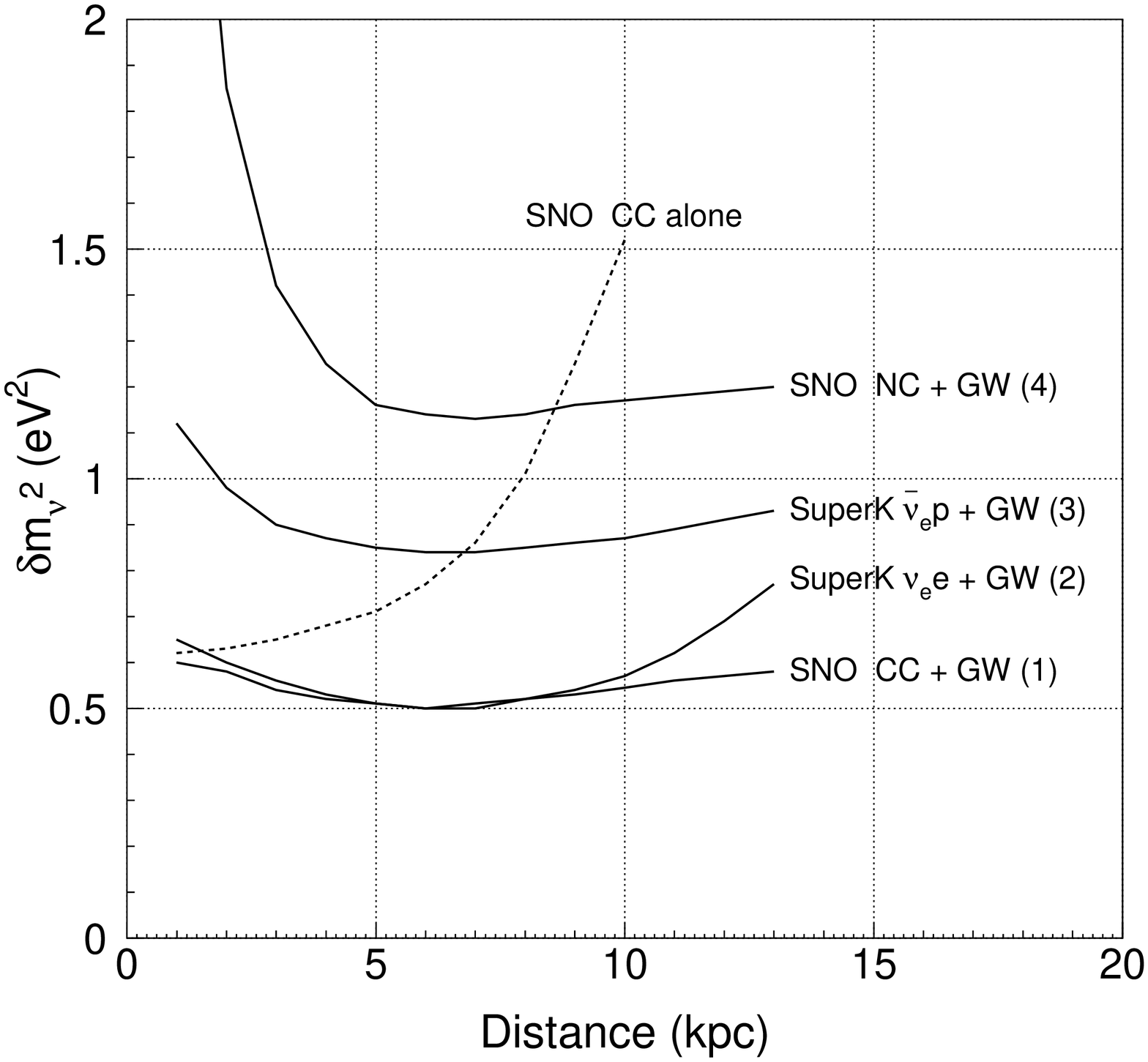,width=20cm}}
Figure~8: Comparison of the sensitivities on the neutrino mass squared for the four methods described in the core of the article. Sensitivities are plotted versus the supernova distance (in kpc). In addition, a fifth curve (dashed line) shows the sensitivity that the SNO neutrino detector could achieve by itself.
\end{figure}

\begin{figure}[here!]
\centerline{\epsfig{file=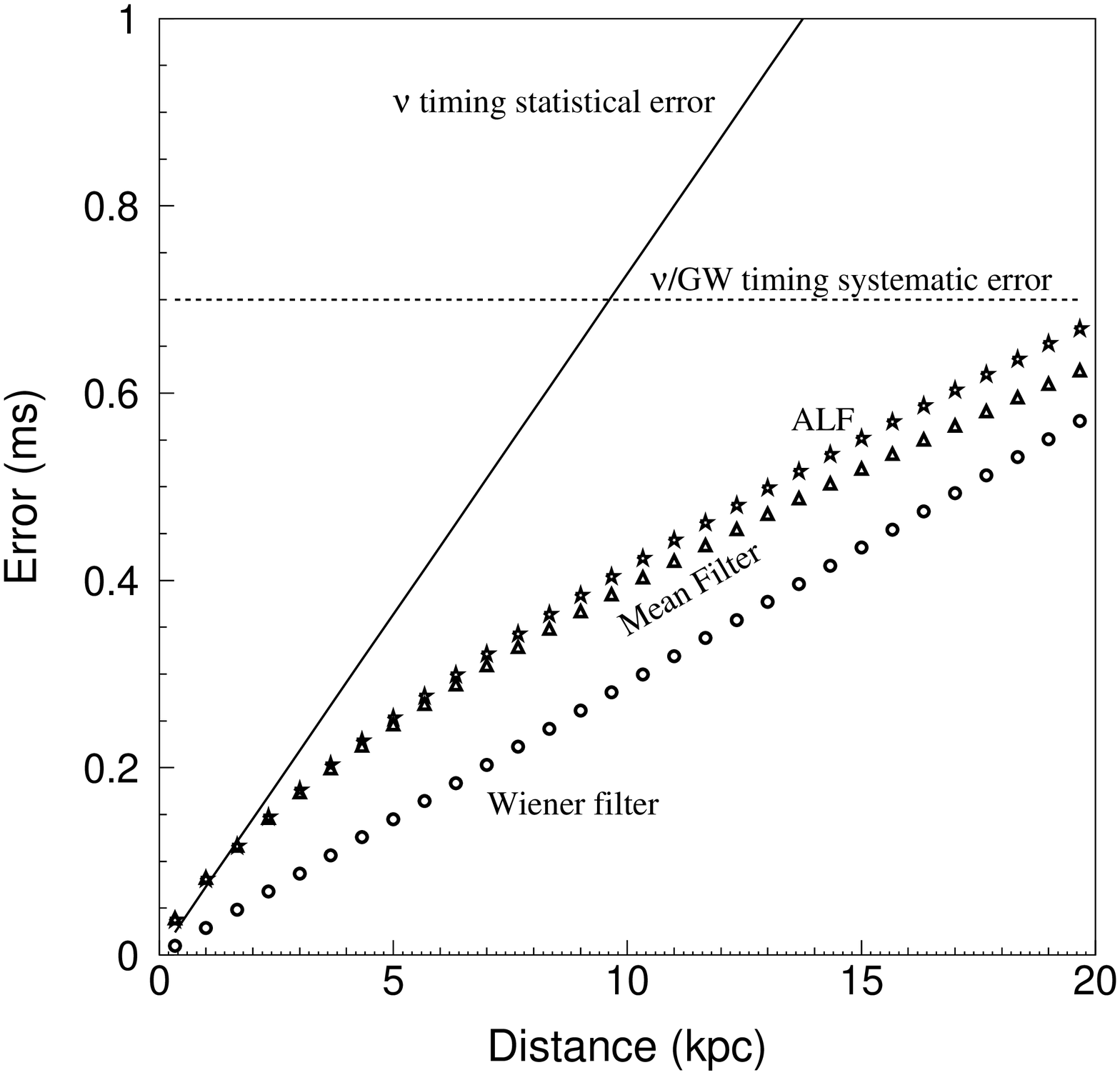,width=20cm}}
Figure~9: Evolution of the different error contributions to the measurement of the delay GW-$\nu$ versus the supernova distance (in kpc). The statistical errors are computed using Eq. (\ref{eq:nu_stat}) and (\ref{eq:GW_stat}), assuming that $N_{\mathrm{e}}=10$ $(N_{\mathrm{e}}\propto D^{-2})$ neutrinos of a supernova exploding at $D=10$~kpc are detected and that its GW countepart is a millisecond burst of SNR $\rho=5$ $(\rho \propto D^{-1})$.
\end{figure}

\begin{figure}[here!]
\centerline{\epsfig{file=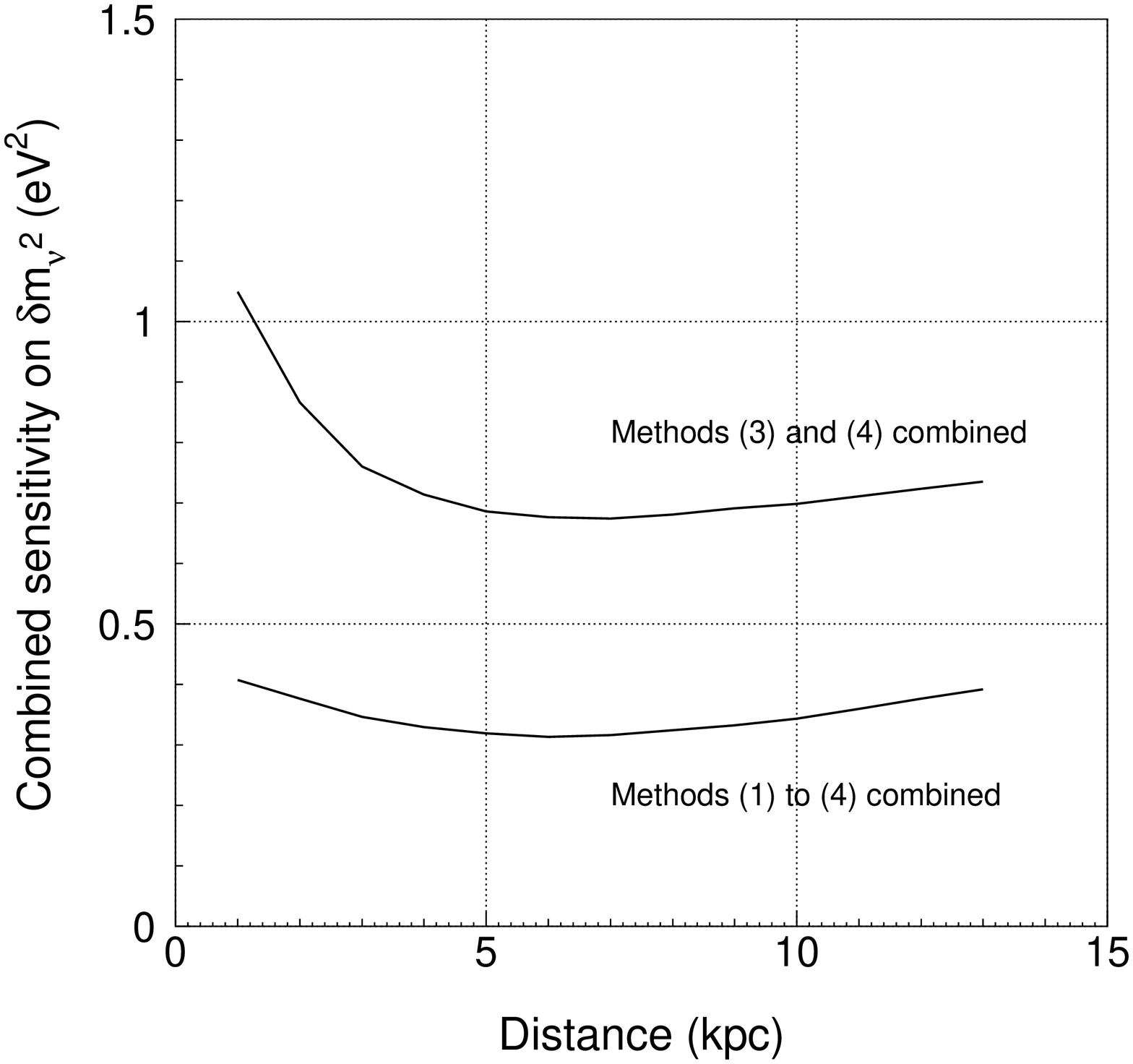,width=20cm}}
Figure~10: Combined sensitivity on the neutrino mass squared versus the supernova distance (in kpc). If $\nu_{\mathrm{e}}$ survive -- $P_{\mathrm{e}}=0.5$ -- the four methods can be combined, whereas only methods (3) and (4) can be used if the $\nu_{\mathrm{e}}$ vanish -- $P_{\mathrm{e}} \sim 0$.
\end{figure}

\end{document}